\documentstyle[aps,preprint,eqsecnum]{revtex}
\begin{document}

\newcommand{\be}{\begin{equation}}
\newcommand{\ee}{\end{equation}}
\newcommand{\bea}{\begin{eqnarray}}
\newcommand{\eea}{\end{eqnarray}}
\newcommand{\nn}{\nonumber}

\setcounter{page}{0}
\draft
\preprint{
\begin{minipage}{5cm}
MPG-VT-UR 109/97\\
E2-97-217
\end{minipage}}

\title{Hamiltonian reduction of $SU(2)$ Dirac-Yang-Mills mechanics}
\author{ S.A. Gogilidze  \ $^a$, A M. Khvedelidze \ $^a$
\thanks{Permanent address: Tbilisi Mathematical Institute,
380093, Tbilisi, Georgia}, \,\,  D.M.Mladenov \ $^a$,
\,\, H.-P. Pavel\ $^b$ }
\address{$^a$ Bogoliubov Theoretical Laboratory, Joint Institute for Nuclear
Research, Dubna, Russia}
\address{$^b$ Fachbereich Physik der Universit\"at Rostock,
              D-18051 Rostock, Germany}
\date{\today}
\maketitle
\begin{abstract}
The $SU(2)$ gauge invariant Dirac-Yang-Mills mechanics of
spatially homogeneous isospinor and gauge fields is
considered in the framework of the generalized Hamiltonian approach.
The unconstrained Hamiltonian system equivalent to the model is obtained
using the gaugeless method of Hamiltonian reduction.
The latter includes the Abelianization of the first class constraints,
putting the second class constraints into the canonical form and
performing a canonical transformation to a set of adapted coordinates such
that a subset of the new canonical pairs coincides with the second class
constraints and part of the new momenta is equal to the Abelian constraints.
In the adapted basis the pure gauge degrees of freedom
automatically drop out from the consideration
after projection of the model onto the constraint shell.
Apart from the elimination of these ignorable degrees of freedom a
further Hamiltonian reduction is achieved due to the three dimensional
group of rigid symmetry possessed by the system.
\end{abstract}

\bigskip


\bigskip
\bigskip

\section{Introduction}

The correct canonical formulation of the quantum theory of non-Abelian fields
assumes a detailed  knowledge of the corresponding
classical generalized Hamiltonian dynamics \cite{DiracL}-\cite{HenTeit}.
Since the introduction of non-Abelian gauge fields by C.N. Yang and
R.L. Mills \cite{YM} over forty years ago
essential progress in this direction has been made.
Rigorous statements about the geometrical structure of the configuration
and the phase space have been established.
It turned out that due to the underlying non-Abelian gauge
symmetry  the ``true phase space'' of Yang-Mills theory, namely
the quotient space of phase space by the action of gauge transformations,
possesses a rich topological structure \cite{Geom}.
In the framework of traditional perturbation theory these geometrical
peculiarities are not taken into account and as a result
the description of large scale effects, including confinement,
is beyond its scope.
The most important lesson one has learned is that, in order to reach a
complete description,
it is necessary to first reformulate Yang-Mills theory in terms of
gauge-invariant variables and only after this step apply
any approximation method.
With this aim several different representations for the
physical degrees of freedom of non-Abelian theories
\cite{GoldJack}-\cite{KhZer0} have been proposed.
All these approaches lead to an unconstrained  Hamiltonian
system, which exhibits non-perturbative features
and are in some sense alternatives to the
conventional perturbative approach. Whereas perturbation theory is appropriate
for the computation of short distance effects, the unconstrained
formulation is adapted to the study of large scale phenomena if the gauge
invariant expressions are evaluated in a derivative expansion.
Since the work by Matinyan et al. \cite{MatSav},
the corresponding zeroth order or
long-wavelength approximation, the Yang-Mills mechanics of
spatially homogeneous gauge fields,
has been studied extensively from different points of view
(see e.g. \cite{AsatSav} - \cite{DahRaab} and references therein).
In the present note we shall continue the study of the model
arising in this approximation, pursuing the aim to prepare the necessary
background for studying the problem of construction of the reduced
phase space of QCD.
Due to the spatial homogeneity condition conventional
Dirac-Yang-Mills theory reduces to a theory describing a
finite dimensional system
which is incomparably simpler than the exact field system.
At the same time, however, it possesses all the main
peculiarities of the full theory and can be used as a laboratory for
testing the viability of ideas and techniques that could be applied
in the general case.

Below we shall isolate the true dynamical degrees of freedom of
$SU(2)$ Dirac-Yang-Mills theory in the long-wavelength approximation using
the gaugeless approach
\footnote{Presumably, S.Shanmugadhasan \cite{Shanmugadhasan}
was the first to employ the classical Lee-Cartan method
of reduction (see e.g.\cite{Cartan} - \cite{Perelomov})
in the framework of generalized Hamiltonian dynamics.}
to the reduction in the number of degrees of freedom
instead of the conventional gauge fixing method.
\footnote{We point out here that the idea of constructing the
physical variables entirely in internal terms without using any additional
gauge conditions is connected
with the desire not to distort the global properties of the theory
and to have all dynamical degrees of freedom under control.}
The cornerstones for this method applied to a system
with first class constraints are the procedure of Abelianization of
constraints (replacement of the
original non-Abelian constraints by an equivalent set of Abelian ones)
and the canonical transformation to new variables where a subset of the
new momenta is equal to the new Abelian constraints.
The system of interacting gauge and spinor
fields considered in this article represent a Hamiltonian system
with mixed first and second class constraints. In this case
the reduction procedure additionally includes the separation of first and
second class constraints and putting them into the canonical form.

The paper is organized as follows.
In Section II we briefly recall how to obtain the unconstrained Hamiltonian
system from the initially gauge symmetric one in the framework of Dirac
constraint theory in order to set the formalism.
The Dirac and the Faddeev gauge
fixing methods as well as the gaugeless method are described.
In Section III the gaugeless method is exemplified by considering the
Yang-Mills system in \(0+1\)- dimensions.
In Section IV we perform the reduction of the Dirac-Yang-Mills
system by explicitly
separating the first and second class constraints, putting the second class
constraints into the canonical form and
Abelianizing the first class constraints.
We construct the corresponding reduced Hamiltonian system
by first eliminating the unphysical gauge degrees of freedom
and then
using the classical scheme of Hamiltonian reduction due to the existence
of three first integrals of motion.
Section V finally gives our conclusions and remarks.

\section{ Reduction of constrained systems with first class constraints}

\label{sec:RFC}

The procedure of reduction of phase space of a singular system
is a generalization of the method of reduction of a system
of differential equations possessing a Lie group symmetry.
The well-known results for this type of reduction in the number of the
degrees of freedom are embodied in the famous
J.Liouville theorem on first integrals in involution.
Interest in these has revived in connection with the study
of Hamiltonian systems with a local (gauge) symmetry.
Since the work of P. Bergmann and P.A.M. Dirac at the beginning of
the fifties it has become clear that the role of integrals of motion in
a Hamiltonian system with gauge symmetry is played by the first class
constraints. Although the reduction
in the number of degrees of freedom due to first class constraints
has many features in common with the classical case, there are
very important differences.
In order to explain these peculiarities of the reduction procedure
and to make the paper self-contained we first have to summarize some
definitions and to put facts from the Dirac theory of generalized
Hamiltonian dynamics into the appropriate context.
In view of the main purpose of our paper, namely to study the finite
dimensional system of homogenous Yang-Mills fields,
we shall discuss the above ideas
for a mechanical system, i.e. a system with a finite number
of degrees of freedom.

\subsection{The definition of reduced phase space}

Let us consider a system with the \( 2n \) - dimensional
Euclidean phase space \(\Gamma \) spanned by the canonical coordinates
\( q_i \) and their conjugate momenta
\( p_i\) and  endowed with the canonical
simplectic structure \(\{ q_i, p^j\}= \delta_i^j \).
Suppose that the dynamics is constrained to
a certain \( (2n - m) \) --- dimensional submanifold
\(\Gamma_c \) determined by \( m \)
functionally independent constraints
\be  \label{eq:constr}
\varphi_\alpha (p,q) = 0~,
\ee
which we assume to be first class
\be \label{eq:fconstr}
\{ \varphi_\alpha (p,q), \varphi_\beta  (p,q)\} =
f_{ \alpha \beta \gamma} (p,q) \varphi_\gamma  (p,q)
\ee
and complete in the sense that
\be \label{eq:comp}
\{ \varphi_\alpha (p,q), H_C (p,q)\} =
g_{\alpha\gamma}\varphi_\gamma (p,q)~,
\ee
where
\(H_C (p,q) \) is the  canonical Hamiltonian.
Due to the presence of these constraints the Hamiltonian system admits
generalized dynamics described by the extended Poincare-Cartan form
\be  \label{eq:P-C}
\Theta : =  \sum_{i=1}^{n} \, p_i dq_i -  H_E (p,q) dt
\ee
with the extended
Hamiltonian \(H_E (p,q) \) that differs from the canonical
\( H_C (p,q)\)  by  a linear combination of  constraints with arbitrary
multipliers \( u_\alpha (t) \)
\be
H_E (p,q) : = H_C (p,q) + u_\alpha (t) \varphi_\alpha (p, q)~.
\ee
>From the condition of completeness (\ref{eq:comp}) with \(H_C\)
replaced by \(H_E\) it follows that
for first class constraints the functions
\( u_\alpha (t) \) can not be fixed in internal terms of the theory.
This implies that the system possesses a local  symmetry
and  that the coordinates split up into two sets, one set whose dynamics
is governed in an arbitrary way and another set with an uniquely determined
behaviour.
Recalling the Dirac definition \cite{Dirac49} of a
{\it physical variable}
as  a dynamical variable \(  F \) with the property
\be                \label{eq:physvar}
     \{ F (p,q ), \varphi_\alpha (p,q) \}  =
d_{ \alpha \gamma} (p,q) \varphi_\gamma  (p,q)~,
\ee
one can conclude that the first set of coordinates does not affect the
physical quantities which are defined on some subspace of the
constraint surface \( {\Gamma}_{c}\).
Indeed, if one considers (\ref{eq:physvar})
as a set of \( m \) first order linear differential equations for
\( F \), then due to the integrability condition (\ref{eq:fconstr})
this function  can be  completely determined by its values in the
\( 2(n-m) \) submanifold of its initial conditions
~\cite{Faddeev69}, \cite{KonPop}.
This subspace of constraint shell represents the
{\it reduced phase space}
\( {\Gamma}^{\ast} \).  This definition of reduced phase space is
implicit. The  main problem is to find the set of \( 2(n-m) \)
`` physical coordinates'' \(Q^{\ast}_i,\: P^{\ast}_i \)
that span this reduced phase space and pick out the
other additional \( m \) pairs which
have no physical significance and represent the pure gauge degrees of
freedom.
Several approaches to its solution are known.
Below we shall briefly discuss the corresponding methods of practical
construction of the physical and the gauge degrees of freedom
with and without gauge fixing.

\subsection{Reduced phase space with the Dirac gauge fixing method}

General principles for imposing gauge fixing constraints
onto the  canonical variables in the Hamiltonian approach were proposed by
Dirac in connection with the canonical formulation of gravity \cite{Dirac59}.
According to the Dirac gauge fixing prescription, one starts
with the introduction of as many new ``gauge'' constraints
\be \label{eq:gauge2}
\chi_\alpha (p,q) = 0
\ee
as there are first class constraints (\ref{eq:constr}),
with the requirement
\be \label{eq:regfp}
\det \Vert \{\chi_\alpha (p,q), \varphi_\beta (p,q)\} \Vert \, \not= \, 0~.
\ee
This allows one to find the unknown Lagrange multipliers
\(u_\alpha(t)\) from the requirement of
conservation of the gauge conditions (\ref{eq:gauge2}) in time
\footnote{Everywhere in the article the dot over the letter
denotes the derivative with respect to the time variable}
\be
 \dot{\chi}_\alpha = \{\chi_\alpha, H_C \} + \sum_{\beta} \{\chi_\alpha,
\varphi_\beta  \} u_\beta = 0
\ee
and thus to determine the dynamics of system in a unique manner.
A striking result of Dirac consists in the observation
that such kind of fixation of Lagrange multipliers \(u(t)\)
is equivalent to the following way of proceeding.
One can drop both the constraints
(\ref{eq:constr}) and the gauge fixing conditions
(\ref{eq:gauge2})
and at the same time achieve the
reduction to the unconstrained theory by using
the Dirac brackets
\be  \label{eq:DB}
\{F , G \}{}_{D} : =  \{F, G \} -
\{F ,\xi _s\}C^{-1}_{s s'} \{ \xi_{s'}, G \}~,
\ee
instead of the Poisson brackets. Here \(\xi\) denotes the set of
all constraints (\ref{eq:constr}) and (\ref{eq:gauge2})
and \( C^{-1} \) is the inverse of the Poisson matrix
\(~
C_{\alpha \beta} : = \{\xi_\alpha , \xi_\beta\}~.
\)
In this method all coordinates of the phase space are treated on an equal
footing
and all information on both initial and gauge constraints is absorbed into
the Dirac brackets, which describe the effective reduction in the number of
degrees of freedom from $n$ to $n-m$
\[
\sum_{i =1}^{n}
\{q_i, p_i, \}_{P.B.} = n~, \quad
\sum_{i =1}^{n}
\{q_i, p_i, \}_{D.B.}
= {n-m}~.
\]
The inclusion of  gauge constraints in addition to the initial
constraints allows one  to take the constraint nature of the
canonical variables into account  by changing the
initial canonical symplectic structure to a new one defined by the Dirac
brackets. The new canonical structure, being dependent on the choice
of gauge fixing-conditions, is  very complicated in general and
it is not clear how to deal with it, in particular, when we are quantizing
the theory.
However, there is a special case when
the Dirac bracket coincides with the canonical one
and looks like the Poisson bracket for an unconstrained system
defined on \({\Gamma}^{\ast}\)
\be  \label{eq:PDB}
\{F,G \}_{D}
\Bigl\vert_{ \varphi = 0,~\chi = 0~}
=  \sum_{i=1}^{n-m}\left\{
\frac{\partial \overline{F}}{\partial Q^{\ast}_i}
\frac{\partial \overline{G}}{ P^{\ast}_i} -
\frac{\partial \overline{F}}{\partial P^{\ast}_i}
\frac{\partial \overline{G}}{ Q^{\ast}_i}\right\}.
\ee
This representation of the Dirac bracket means that
in terms of the conjugate  coordinates \(Q^{\ast}_i, P^{\ast}_i
\, (i = 1, \dots , n-m ) \)  the reduced phase space
is parametrized so that constraints  vanish identically and any
function \( F(p,q) \) given on the
reduced phase space becomes \cite{Sunder}
\[
 F(p,q)
\Bigl\vert_{ \varphi = 0, ~\chi = 0~} \, = \overline{F}(P^{\ast}, Q^{\ast})
~.\]
Thus in the Dirac gauge-fixing method the problem of definition of
the ``true dynamical  degrees'' of freedom reduces to the
problem of a ``lucky'' choice of the gauge condition.

\subsection{Reduced phase space with the Faddeev gauge fixing method}

An alternative to the  Dirac gauge-fixing procedure has been proposed
in the well-known paper by L.D. Faddeev \cite{Faddeev69},
devoted to the method of path integral quantization of a constrained
system. In contrast to the Dirac method, the main idea of the
Faddeev method is to introduce an explicit parametrization of
the reduced phase space.
As in the Dirac method, one introduces  gauge fixing constraints
\(
\chi_\alpha (p,q) = 0~,
\)
but now with the additional `` Abelian'' property
\be                \label{eq:abg}
     \{ \chi_\alpha (p,q), \chi_\beta(p,q) \}  =  0~,
\ee
and the requirement  (\ref{eq:regfp})
is fulfilled.
Now, in accordance with the Abelian character of
gauge conditions (\ref{eq:abg}), there exists a canonical
transformation to new coordinates
\bea
q_i & \mapsto & Q_i : = Q_i \left ( q , p \right )\nn\\
p_i & \mapsto & P_i : = P_i \left ( q , p \right )
\eea
such  that  \( m \) of the new  \( P  \) 's
coincide with the constraints $\chi_{\alpha}$
\be
{P}_\alpha = \chi _\alpha \left ( q , p \right ).
\ee
The condition  (\ref{eq:regfp}) allows one to resolve
the constraints (\ref{eq:constr})
for the coordinates \( Q_\alpha\)
in terms of the  \( (n - m )\) canonical pairs
(\({Q}^{\ast}_i, {P}^{\ast}_i\)), which span the
\(2(n-m) \) - dimensional surface \(\Sigma\)
determined by the equations
\bea
P_\alpha & = & 0~, \nn\\
Q_\alpha & = & Q_\alpha \left ( {Q}^{\ast}, {P}^{\ast}  \right)~.
\eea
After this construction has been carried out, the problem is to prove that
the surface \(\Sigma\) coincides with the true reduced phase space
\({\Gamma}^{\ast} \),  independent of the choice of
the gauge fixing conditions.
In other words, it is necessary to find a criterion for gauge conditions
to be admissible.
A radical method to solve this problem is
not to use any gauge conditions at all.
The following subsection will give a brief description of such an
alternative gaugeless scheme to construct the reduced phase space without
using gauge fixing functions.

\subsection{The gaugeless method }
\label{subsec:glr}

If the theory contains only Abelian constraints one can find a
parametrization of reduced phase space as follows.
According to a well-known theorem (see e.g. \cite{Whittaker}),
it is always possible to find a canonical transformation to a new
set of canonical coordinates
\bea \label{eq:cantr}
q_i & \mapsto & Q_i : = Q_i \left ( q , p \right ),\nn\\
p_i & \mapsto & P_i : = P_i \left ( q , p \right ),
\eea
such  that  \( m \) of the new  momenta,
(\(\overline{P}_1, \dots ,\overline{P}_m \)),
become equal to the Abelian constraints $\varphi_{\alpha}$
\be
\overline{P}_\alpha = \varphi _\alpha (q, p)~.
\ee
In terms of the new coordinates (\(\overline{Q}, \overline{P}\)), and
(\( {Q}^{\ast}, {P}^{\ast}\)) the canonical equations read
\bea \label{eq:motion}
 \dot{Q}^{\ast} = \{{Q}^{\ast}, H_{phys} \}~, &\quad \quad  \quad &
\dot{\overline{Q}} = u(t)~, \nn\\
 \dot{P}^{\ast} = \{{P}^{\ast}, H_{phys} \}~, & \quad \quad  \quad &
\dot{\overline{P}} = 0~,
\eea
with the physical Hamiltonian
\be \label{eq:PhysHam}
 H_{phys}(P^{\ast}, Q^{\ast}) \equiv H_C(P, Q)\,
 \Bigl\vert_{ \overline{P}_\alpha = 0}.
\ee
\(H_{phys}\) depends  only on the \(( n-m )\)
pairs of new gauge invariant canonical coordinates
\(( {Q}^{\ast}, {P}^{\ast} )\) and the  form of the canonical system
(\ref{eq:motion}) expresses the  explicit separation of the phase space into
physical and unphysical sectors
\bea
2n \left \{ \left (
\begin{array}{c}
q_1\\
p _1\\
\vdots\\
q_n\\
p _n\\
\end{array}
\right )  \right.
\quad    \mapsto  \quad
\begin{array}[c]{r}
{ 2(n-m)\,\, \left\{ \left(
\begin{array}{c}
Q^\ast \\
P^\ast
\end{array}
\right ) \right. }\\
\vphantom{\vdots} \\
{ 2m\, \left\{ \left(
\begin{array}{c}
\overline{Q}\\
\overline{P}
\end{array}
\right ) \right. }
\end{array}
\quad
\begin{array}[@]{c}
{ \begin{array}{c}
{\it physical} \\
{\it variables}
\end{array}
} \\
\vphantom{\vdots} \\
{ \begin{array}{c}
{\it unphysical}\\
{\it variables }
\end{array}}
\end{array}
\eea
The arbitrary functions \( u(t) \) enter into that part of the system of
equations, which contains only the ignorable coordinates
\( \overline{Q}_\alpha \) and momenta \({\overline{P}}_\alpha\).
A straightforward generalization of this method to the non-Abelian case is
not possible, since the identification of momenta with constraints is
forbidden due to the non-Abelian character of the constraints.
However, there exists the  possibility  of
a  replacement of  the constraints $\varphi_{\alpha}$ by an equivalent set of
new constraints $\Phi_{\alpha}$
\be  \label{eq:abelianiz}
\Phi_\alpha = D_{\alpha \beta}\varphi_\beta\,,   \quad \quad
\det \Vert {D} \Vert \:
\Bigl\vert_{ \varphi = 0 }\: \not=  \,0~,
\ee
describing the same surface \(\Gamma_c \) but  forming  an Abelian algebra.
There are different proofs of this statement,
based on the resolution of constraints
\cite{Sunder} -- \cite{HenTeit}, exploiting gauge-fixing conditions
\cite{BatVil}, or using the direct  method   of constructing the
Abelianization matrix as the solution of a certain system of linear first
order differential equations  \cite{GKP}
\footnote{In all cases, the proofs use the
large freedom in the canonical description of the constrained systems.
Apart from the ordinary canonical transformations there exist
generalized canonical transformations \cite{Bergman} i.e.,
those which preserve the form of all constraints
of the  theory as well as the canonical form of the equations of motion.
The Abelianization  transformation (\ref{eq:abelianiz}) is of
course non-canonical, but belongs to this class of
generalized canonical transformations.}.
For non-Abelian systems therefore, the construction of the Abelianization
matrix and
the implementation of the above mentioned transformation
(\ref{eq:cantr}) to the new set of Abelian constraint functions
$\Phi_{\alpha}$ completes the reduction of the phase space without
using gauge fixing functions, solely in internal terms of the theory.

Before applying the gaugeless method to the construction  of the reduced
phase space of homogeneous Yang-Mills fields in 3+1-dimensional space
it seems worth setting forth our approach to the same problem
in 0+1-dimensional space.

\section{$ SU(2)$ Yang-Mills fields in 0+1 dimensions }
\label{sec:YM01}

In order to explain our main idea how to construct the physical variables
we shall start with the non-Abelian Christ $\&$ Lee model
\cite{ChrLee}, \cite{Prokh}.
The Lagrangian of this model is
\begin{equation}
L : = \frac{1}{2}\ (D_t x)_i (D_t x)_i -  \frac{1}{2}\ V( x^2 )~,
\end{equation}
where \(x_i\) and \(y_i\) are the components of  three-dimensional vectors
and the covariant derivative  $D_t$ is defined as
\be
(D_t x)_i := \dot{x}_i + g \epsilon_{ijk} y_j x_k~.
\ee
One can see that this model is nothing else than Yang-Mills
theory in $0+1$ dimensional space-time
and that is invariant under \(S0(3)\) gauge transformations.

Performing the Legendre transformations
\bea
p_y^i &= & \frac{\partial L}{\partial \dot{y_i}}~, \\
p^{i} & = & \frac{\partial L}{\partial \dot{x_i}} =
\dot{x_{i}} + g \epsilon^{ijk} y_j x_{k}~,
\eea
one obtains the canonical Hamiltonian
\be \label{eq:khrlh}
H_C = \frac{1}{2} p_{i} p_{i}
- \epsilon_{ijk} x_{j} p_{k} y_i  +  V(x^2)~,
\ee
and identifies the three primary constraints
\(
\,{p_y^i} =  0 \,
\)
as well as the three secondary ones
\bea \label{eq:secchr}
\Phi_i  = \epsilon_{ijk} x_j p_k = 0~,
\eea
obeying the \(SO(3)\) algebra
\bea \label{eq:SO(3)alg}
\{\Phi_i , \Phi_j \} &= & \epsilon_{ijk}\Phi_j~.
\eea
One easily verifies that  the secondary constraints are functionally dependent,
\(x_i\Phi_i = 0\). We shall now carry out the Abelianization procedure
and choose
\be
\Phi^{(0)}_1  : =  x_2 p_3 -x_3 p_2~, \,\,\,\, \,\,
\Phi^{(0)}_2  : =  x_3 p_1 -x_1 p_3~,
\ee
as the two independent constraints with the algebra
\be \label{eq:algchl}
\{\Phi^{(0)}_1 ,\Phi^{(0)}_2 \} =
- \frac{x_1}{x_3}\, \Phi^{(0)}_1 - \frac{x_2}{x_3}\, \Phi^{(0)}_2~.
\ee
The general iterative scheme of the construction of Abelianization matrix
\cite{GKHAbel} consists of two steps for this simple case.
Let us at first exclude \( \Phi^{(0)}_1\) from the right hand side of eq.
(\ref{eq:algchl}).
This can be achieved by performing the transformation
\bea
\Phi^{(1)}_1  & : =  & \Phi^{(0)}_1~, \nn\\
\Phi^{(1)}_2  & : =  & \Phi^{(0)}_2  + C\, \Phi^{(0)}_1~,
\eea
with the function \(C\) obeying the partial differential equation
\bea
\{\Phi^{(0)}_1 , C \} & = &
 -\frac{x_2}{x_3}\, \, C + \frac{x_1}{x_3}~.
\eea
Writing down a particular solution of this equation
\be
C(x) =  \frac{x_1 x_2}{x_2^2 + x_3^2}~,
\ee
we get the algebra for new constraints
\bea
\{\Phi^{(1)}_1 ,\Phi^{(1)}_2 \} &= &
- \frac{x_2}{x_3}\, \Phi^{(1)}_2~.
\eea
Now let us perform the second transformation
\bea
\Phi^{(2)}_1  &  : =   & \Phi^{(1)}_1 ~,\nn\\
\Phi^{(2)}_2  &  : =   & B\, \Phi^{(1)}_2~,
\eea
with the function \(B \)  satisfying the equation
\bea
\{\Phi^{(2)}_1 ,B \} & = & \frac{x_2}{x_3}\, B~.
\eea
A particular  solution of  this equation is
~\(
B(x) =  \frac{1}{x_3}~.
\)
As result of the two above transformations, the Abelian constraints
equivalent to the initial non-Abelian ones
have the form
\bea \label{eq:Chleeab}
\Phi^{(2)}_1 & = & x_2 p_3 - x_3 p_2\,, \nn \\
\Phi^{(2)}_2  & = & \frac{1}{x_3} \left[ (x_3 p_1 -x_1 p_3)
+ \frac{x_1 x_2}{x_2^2 + x_3^2}
( x_2 p_3 - x_3 p_2 )
\right] \,.
\eea

\subsection{ Canonical transformation and reduced Hamiltonian}

We are now ready to perform a canonical transformation
to new variables so that two new momenta
will coincide with the Abelian constraints (\ref{eq:Chleeab})
\footnote{Here we introduce the compact notations for three-dimensional vectors
\(\vec x ,  \vec p \) and multiply the constraint \(\Phi^{(2)}_2 \) by the
factor
\( \sqrt{{x_2}^2 + {x_3}^2}\)  to deal with constraints of one and the
same dimension. This multiplication conserves the Abelian character of
the constraints, since
\( \{ \Phi_1^{(2)} , \sqrt{{x_2}^2 + {x_3}^2} \} = 0 \).}
\be
\label{eq:pthetapphi}
p_\theta  :=  \frac{ (\vec x \cdot \vec p) \, x_1 - {\vec x}^2 \, p_1}
{\sqrt{{x_2}^2 + {x_3}^2}}~, \,\,\,\,\,\,\,
p_\phi :=  x_2p_3 -x_3p_2~.
\ee
It is easy to verify that the contact transformation from
the Cartesian coordinates to the spherical ones
\bea
x_1 = r \cos\theta,~~~~~~ & \quad \quad & r = \sqrt{{x_1}^2 + {x_2}^2 + {x_3}^2
},\nn \\
x_2 = r \sin\phi \sin{\theta}, & \quad \quad &
\theta = \arccos \frac{x_1}{\sqrt{{x_1}^2 + {x_2}^2 + {x_3}^2 } }, \nn  \\
 x_3 = r \cos\phi \sin\theta,  & \quad \quad &   \phi  =
\arctan\,
\left(\frac{x_2}{x_3}\right)~,
\label{eq:rphitheta}
\eea
is just the required transformation.
Indeed, using the corresponding generating function
\be
F [\vec x ;\ p_r, p_\theta, p_\phi ] =
p_r \sqrt{{x_1}^2 + {x_2}^2 + {x_3}^2 } +
p_{\theta} \arccos \frac{x_1}{\sqrt{{x_1}^2 + {x_2}^2 + {x_3}^2 } }
+ p_{\phi} \arctan\,
\left(\frac{x_2}{x_3}\right)~,
\ee
we get
\bea
p_1 & = & \frac{\partial F}{\partial x_1} =
p_{r} \cos\theta - p_{\theta} \frac{\sin\theta}{r}~, \\
p_2 & = & \frac{\partial F}{\partial x_2} =
p_{r} \sin\theta \sin\phi + p_{\theta} \frac{\sin\phi \cos \theta}{r}
+ p_\phi \frac{\cos \phi}{r \sin\theta}~, \\
p_3 & = & \frac{\partial F}{\partial x_3} =
p_{r} \sin\theta \cos\phi + p_{\theta} \frac{\cos\phi \cos \theta}{r}
- p_\phi \frac{\sin\phi}{r \sin\theta}~,
\eea
and convince ourselves that in terms of these new variables the two
independent constraints are indeed $p_\theta=0$ and $p_\phi=0$
in accordance with (\ref{eq:pthetapphi}).
It is worth noting here that starting with the set of reducible
constraints (\ref{eq:secchr})  and performing the above transformation
(\ref{eq:rphitheta}) one obtains the representation
\bea
\Phi_1 & = &  -p_\phi\,,\\
\Phi_2 & = &  -p_\theta \cos \phi + p_\phi \sin \phi \cot \theta\,,\\
\Phi_3 & = &  p_\theta \sin \phi +p_\phi \cos \phi\cot \theta~,
\eea
adapted to the Abelianization.
The corresponding Abelianization matrix for the reducible set of
constraints is
\be
D := \frac{1}{d}\left (
\begin{array}{ccc}
-d_2 \sin\phi -d_3 \cos\phi\,,            &  d_1 \sin\phi\,,
&     d_1 \cos\phi \,,     \\
(d_2 \cos\phi -d_3 \sin\phi)\cot\theta \,, & -d_3 - d_1 \cos\phi\cot\theta \,,
&  d_2 +  d_1 \sin\phi\cot\theta\,,\\
\cot\theta \,,     &   \sin\phi   \,,    &     \cos\phi \,,
\end{array}
\right ),
\ee
with arbitrary
\( \vec d \) and \( d:= d_1 \cot\theta + d_2 \sin\phi + d_3 \cos\phi
\).
This example demonstrates two important features of the Abelianization
procedure:
{\it i)}
it is not necessary to work with an irreducible set of constraints,
because the Abelianization procedure leads automatically to an
irreducible set of
constraints,
{\it ii)} in certain special coordinates the
problem of the solution of differential equations reduces to the
solution of a simple algebraic problem.
In terms of the new canonical variables the canonical Hamiltonian
(\ref{eq:khrlh}) reads
\be  \label{eq:cnhcr}
 H_C =  \frac{1}{2} p^2_r +  \frac{1}{2r^2}\, \left( p^2_\theta +
\frac{p^2_\phi}{\sin^2 \theta}\right)
-p_\phi y_\phi - p_\theta y_\theta  + V(r)~,
\ee
with the physical momentum
\(
p_r = \frac{\left({\vec x}\cdot{\vec p}\right)}{\sqrt{ {x_1}^2 + {x_2}^2 +
{x_3}^2}}~,
\)
and
\bea
y_\phi   &:= & y_1 + y_2  \sin\phi + y_3 \cos\phi\cot\theta \,,\nn\\
y_\theta &:= & y_2 \cos\phi -y_3  \sin\phi\,.
\eea
As a result, all the unphysical variables are separated from
the physical \(r\) and \(p_r\) and
their dynamics is governed by the physical Hamiltonian obtained from
the canonical one by putting  \(p_\phi\) and \(p_\theta\) in
(\ref{eq:cnhcr}) equal to zero
\be
H_{phys}={1\over 2} p_r^2 + V(r)~.
\ee

\section{Spatially homogeneous  $SU(2)$ Dirac-Yang-Mills fields in
3+1 dimensions }

\subsection{Canonical formulation of the model}

The dynamics of \(SU(2)\) Yang-Mills gauge fields \( A^a_{\mu}(x) \)
minimally coupled to the isospinor fields \(\Psi_{\alpha}(x) \)
%
\footnote{The matter isospinor variables \(\Psi_\alpha\)
 are treated classically as
a collection of four Grassmann quantities.
Detailed notations are collected in the Appendix.}
in four-dimensional Minkowski space-time is defined by the
Lagrange density
\be  \label{eq:YML}
{\cal L} = {\cal L}_{Y.-M.} +  {\cal L}_{Matter} + {\cal L}_{I}~.
\ee
The first term is the kinetic term of the non-Abelian fields
\bea
&&{\cal L}_{Y.-M.}  =  \frac{1}{2}\  \mbox{tr}\, \left(
F_{\mu\nu} F^{\mu \nu} \right)~,
\eea
the second term corresponds to the matter part
\be  \label{eq:matter}
{\cal L}_{Matter} = \frac{i}{2} [\bar{\Psi}_\alpha \gamma_\mu
{\partial}^{\mu} \Psi_\alpha -  {\partial}^{\mu} \bar{\Psi}_\alpha
\gamma_\mu \Psi_\alpha ]\,
- m \,\bar{\Psi}_\alpha \Psi_\alpha~,
\ee
and the last term describes the interaction between the gauge and the
matter fields
\be  \label{eq:matterint}
{\cal L}_{I} =  g\ \frac{1}{2}
\bar{\Psi}_\alpha \gamma^\mu ( \tau^a )_{\alpha\beta}\Psi_\beta~A^a_\mu~,
\ee
with the Pauli matrices \( \tau_a,~a=1,2,3 \).

After the supposition of the spatial homogeneity of the fields,
(\ref{eq:YML}) reduces to a finite dimensional model
described by the Lagrangian
\begin{equation} \label{eq:hyml}
L = \frac{1}{2} ( D_t A )_{ai} ( D_t A)_{ai}
+
\frac{i}{2} \left({\bar\Psi}_\alpha \gamma_0\dot{ \Psi}_\alpha -
\dot{{\bar\Psi}}_\alpha \gamma_0 {\Psi}_\alpha \right)
- m \bar{\Psi}_\alpha \Psi_\alpha
-
g \rho_a\, Y_a + g j_{ia} A_{ai} - V(A)~,
\end{equation}
where the nine spatial components
 \( A^{a}_{i} \) are written in the form of a \( 3 \times 3 \) matrix
\( A_{ai}\),
the time component of the gauge potential is identified with
 $ Y_a : =  A_0^a $ and $D_t$ denotes the covariant derivative
$$ (D_t A)_{ai} := \dot{A}_{ai} -  g \epsilon_{abc} Y_b A_{ci}~. $$
The part of the Lagrangian density corresponding to the selfinteraction
of the gauge fields is gathered in the ``potential'' $V(A)$
\be
V(A) : = \frac{g^2}{4}
\left[
 \mbox{tr}{}^2 ( A A^T ) -
 \mbox{tr} ( A A^T )^2
\right]~,
\ee
while their interactions with the matter fields are via the
isospinor currents
\bea
&&\rho_a \left[ \Psi \right]\
:= \  \frac{1}{2}\bar{\Psi}_\alpha \gamma_0 \left( \tau_a
\right)_{\alpha \beta} \Psi_\beta~, \nn\\
&&
j_{ia} \left[ \Psi \right] \ :=  \ \frac{1}{2} \bar{\Psi}_\alpha \gamma_i
\left(
\tau_a
\right)_{\alpha \beta} \Psi_\beta~.
\eea

After Legendre transformation one obtains the canonical Hamiltonian
\be
H_C = \frac{1}{2} E_{ai} E_{ai} +
 m \bar{\Psi}_\alpha \Psi_\alpha
-  g\, \left( \epsilon_{abc} A_{ci} E_{bi}  - \rho_a \right) Y_a
- g\, j_{ia} A_{ai} +  V(A)~,
\ee
defined on the phase space endowed with the canonical
symplectic structure (see Appendix) and spanned by the
bosonic and fermionic canonical variables
\( (Y_a,  P_{Y_a}) \),  \( (A_{ai}, E_{ai}) \)
and \((\Psi_\alpha, P_{\Psi_\alpha}) \),
\( (\bar{\Psi}_\alpha, P_{\bar{\Psi}_\alpha}) \),
where
\bea \label{eq:canm}
P_{Y_a} & := & \frac{\partial L}{\partial \dot{Y}_a} = 0~, \\
E_{ai} & := & \frac{\partial L}{\partial \dot{A}_{ai}} =
\dot{A}_{ai} - g\epsilon_{abc} Y_b A_{ci}~, \\  \label{eq:ferm1}
P_{\Psi_\alpha}& := & L\frac{\stackrel{\leftarrow}{\partial}}
                       {\partial \dot{\Psi}_{\alpha}}
= -\frac{i}{2} \bar{\Psi}_\alpha \gamma_0~,
\\\label{eq:ferm2}
P_{\bar{\Psi}_\alpha}& := & \frac{ \stackrel{\rightarrow}{\partial}}
{\partial \dot{\bar{\Psi}}_{\alpha}}L =
- \frac{i}{2} \gamma_0 {\Psi}_\alpha~.
\eea
According to the definition of the canonical momenta (\ref{eq:canm}),
(\ref{eq:ferm1}) and (\ref{eq:ferm2})  the phase space is restricted
by the three primary bosonic constraints
\be  \label{eq:boscon}
P^a_Y = 0~,
\ee
and the sixteen Grassmann constraints
\be
\Upsilon^1_\alpha  :=
P_{\Psi_\alpha} + \frac{i}{2} \bar{\Psi}_\alpha \gamma_0~, \,\,\,\,\,\,\,\,
\Upsilon^2_\alpha  :=
P_{\bar{\Psi}_\alpha} + \frac{i}{2} \gamma_0 {\Psi}_\alpha~.
\ee
Thus the evolution of the system is governed by
the total Hamiltonian
\be
\label{eq:Htot}
H_T := H_C + u^a_Y (t) \ P^a_Y +
\Upsilon^1_\alpha \ u^1_{\alpha}(t) +
u^2_{\alpha}(t)\ \Upsilon^2_\alpha~.
\ee
The conservation of bosonic constraints  (\ref{eq:boscon})
in time entails the following
further condition on canonical variables
\be  \label{eq:secconstr}
\dot{P}_{Y_a} = 0 \quad \longrightarrow \quad
\Phi_a : = \epsilon_{abc} A_{ci} E_{bi} - \rho_a \left[ \Psi \right] = 0~,
\ee
which is the non-Abelian Gauss law.
In contrast, the  maintenance of Grassmann constraints
$\Upsilon^1_\alpha$ and $\Upsilon^2_\alpha$ in
time allows to determine the Lagrange multipliers $u^1_{\alpha}(t)$ and
$u^2_{\alpha}(t)$
in the expression (\ref{eq:Htot}) for the total Hamiltonian.
Taking into account the  Poisson brackets of constraints
\bea  \label{eq:algb1}
\{\Phi_i , \Phi_j \} &= & \epsilon_{ijk}\Phi_j~ +
\epsilon_{ijk}\, \rho_k \left[ \Psi \right] ~, \\
\{\Phi_a , \Upsilon^1_\alpha \} &=& - \bar{\Psi}_\beta \gamma_0
\left(\tau_a~\right)_{\beta \alpha}, \\
\{\Phi_a , \Upsilon^2_\alpha \}& = & \  \gamma_0
\left(\tau_a~\right)_{\alpha \beta} \Psi_{\beta}~, \\ \label{eq:algb2}
\{ \Upsilon^1_\alpha, \Upsilon^2_\beta \} & = &
- i \delta_{\alpha \beta} \ \gamma_0~,
\eea
one can convince oneself that no new constraints emerge and hence that
ternary constraints are absent in the theory,
\(~~
\dot{\Phi} \biggl\vert_{{Constraint \ Shell}} = 0~.
\)

To implement the reduction procedure without using
gauge fixing conditions we have to put
the constraints into the canonical form discussed in the next paragraph.


\subsection{Putting the constraints into the canonical form}


\subsubsection{Separation of first and second class constraints}

The  set of the 22 constraints \( {{C}_{\cal A}} : =
( P_Y,  \Phi, \Upsilon ) \) represent a mixed system of
first and second class constraints.
The  Poisson matrix
\(
M_{{\cal A} {\cal B} }  := \{ {C}_{\cal A},  {C}_{\cal B} \}
\)
is degenerate on constraint shell,
\(~
\mbox{rank}\, ||{\cal M } || \, \biggl\vert_{\ C_{\cal{A}} \,= \, 0}  = 16.~
\)
Hence among the constraints there are six first class ones.

In order to perform the reduction procedure let us start with the
separation of the first and second class constraints.
The primary constraints \(P_Y\)
``commute'' with all the other constraints and thus we should  deal
only with constraints \( C'_{\cal{A}}: = ( \Phi, \Upsilon ) \).
The separation of constraints is achieved by a transformation to an
equivalent set of constraints  \(\tilde {C}'_{\cal A} : =
(\tilde{\Phi},  \tilde{\Upsilon}  ) \)
\be
\tilde{C}'_{\cal A} : =  D'_{{\cal A} {\cal B}} \ C'_{\cal{B}}\,,
\ee
so that the first class constraints \( \tilde{\Phi} \)
form the ideal of the algebra
\be
\{\tilde{\Phi} , \tilde{\Upsilon} \} = 0\,, \quad \quad
\{\tilde{\Phi} , \tilde{\Phi} \}\bigl \vert_{\Phi = 0}  = 0~,
\ee
and the pairs of second class constraint
satisfy  the canonical algebra
\be
\{\tilde{\Upsilon}^1_\alpha , \tilde{\Upsilon}^2_\beta \}
=  - \ \delta_{\alpha \beta}~.
\ee
In order to transform the algebra of constraints to the canonical form let us
at first perform the  equivalence transformation
\be
\Phi_a^\prime : = \Phi_a + \Upsilon^1_\alpha
\frac{i}{2}~\left( \tau_a \right)_{\alpha \beta}\Psi_{\beta}
+\frac{i}{2}~\bar{\Psi}_\beta \left(\tau_a\right)_{\beta \alpha}
\Upsilon^2_\alpha~,\\
\ee
on the bosonic constraints \( \Phi_a \) and  the equivalence transformation
\be
\tilde{\Upsilon}^1_\alpha  : =  -i {\Upsilon}^1_\alpha \gamma_0~,\,\,\,\,\,\,
\tilde{\Upsilon}^2_\alpha  : = \ \ \,  {\Upsilon}^2_\alpha~,
\ee
on the Grassmann constraints.
The Poisson brackets of the new constraints
\bea
\{\Phi_a^\prime , \Phi_b^\prime \} &= & \epsilon_{abc}\Phi_c^\prime~, \\
\{\tilde{\Upsilon}^1_\alpha, \Phi_a^\prime \} & = & \ \, \frac{i}{2}\,
\tilde{\Upsilon}^1_\beta \left(\tau_a
\right)_{\beta \alpha}, \\
\{\tilde{\Upsilon}^2_\alpha, \Phi_a^\prime \} & = & - \frac{i}{2}\,
\left(\tau_a
\right)_{\alpha \beta} \tilde{\Upsilon}^2_\beta~, \\ \label{eq:cansec}
\{ \tilde{\Upsilon}^1_\alpha, \tilde{\Upsilon}^2_\beta \} & = &
- i \delta_{\alpha \beta}~,
\eea
show the separation of the first class constraints on the surface
\( {\Upsilon} = 0 \) defined by the  second class constraints.
To achieve this separation on the  whole phase space it is necessary to
apply the additional transformation
\be
\tilde{\Phi}_a := \Phi_a^\prime - \tilde{\Upsilon}^1_\alpha \left(
\tau_a \right)_{\alpha \beta} \tilde{\Upsilon}^2_\beta ~.
\ee
One can verify that the first class constraints form the ideal of
the total set of constraints
\bea
\{\tilde{\Phi}_a , \tilde{\Phi}_b \} &= &
\epsilon_{abc}\tilde{\Phi}_c, \\
\{\tilde{\Upsilon}^{1}_\alpha, \tilde{\Phi}_a \} & = & 0, \\
\{\tilde{\Upsilon}^{2}_\alpha, \tilde{\Phi}_a \} & = & 0
\eea
and the second class constraints obey the canonical algebra
(\ref{eq:cansec}).
The explicit form of the resulting set of constraints is
\bea  \label{eq:constrstsec1}
\tilde{\Upsilon}^1_\alpha &: =  &
-i P_{\Psi_\alpha}\gamma_0 + \frac{1}{2} \,
\Psi = 0 ~, \\ \label{eq:constrstsec2}
\tilde{\Upsilon}^2_\alpha  &: = &
P_{\bar{\Psi}_\alpha}  + \frac{1}{2}\, \gamma_0 \Psi = 0 ~, \\
\label{eq:constrstfir}
\tilde{\Phi}_a \, & : = &  \epsilon_{abc} A_{bi}E_{ci} +
\frac{1}{8}\bar{\Psi} \tau_a \gamma_0 \Psi -
\frac{1}{2}P_{\Psi}\tau_a\gamma_0 P_{\bar{\Psi}} +
\frac{i}{4}\left( P_{\Psi}\tau_a{\Psi} +
\bar{\Psi}\tau_a P_{\bar{\Psi}}\right)=0~.
\eea
In order to implement the reduction due to the second class constraints
(\ref{eq:constrstsec1}) and (\ref{eq:constrstsec2})
let us introduce the new canonical variables
$(Q^\ast_{\Psi_\alpha},\bar{Q}_{\Psi_\alpha})$ and
$(\Pi^\ast_{\Psi_\alpha},\bar{\Pi}_{\Psi_\alpha})$ via
\bea
\Psi_\alpha =: i \gamma_0 \left(Q^\ast_{\Psi_\alpha}  -
\bar{Q}_{\Psi_\alpha} \right) ~,&
\quad \quad &
{\bar{\Psi}}_\alpha =: \Pi^\ast_{\Psi_\alpha}  -
\bar{\Pi}_{\Psi_\alpha}~,  \\
P_{\Psi_\alpha} =: \frac{i}{2}
 \left (\bar{\Pi}_{\Psi_\alpha}  -
{\Pi}^\ast_{\Psi_\alpha}\right)\gamma_0~, &&
P_{\bar{\Psi}_\alpha} =: \frac{1}{2}
 \left (\bar{Q}_{\Psi_\alpha}  +
{Q}^\ast_{\Psi_\alpha}\right)~.
\eea
In terms of the new variables the constraints read
\bea
\tilde{\Upsilon}^1_\alpha  & = &{\bar{\Pi}}_{\Psi_\alpha} = 0\, ,\\
\tilde{\Upsilon}^2_\alpha  & = &\bar{Q}_{\Psi_\alpha} = 0~, \\
\tilde{\Phi}_a        & = &\epsilon_{abc} A_{bi}E_{ci} -
 \frac{i}{2}
{\Pi}^\ast_{\Psi_\alpha} \tau_a
{Q}^\ast_{\Psi_\alpha} = 0~ .
\label{eq:phiQstar}
\eea

\subsubsection{Canonical transformation to adapted coordinates }

The example of the Christ and Lee model in Section \ref{sec:YM01}
shows that the realization of constraints by
Abelianization is immediate if one performs a
canonical transformation to a new set of variables containing
the gauge invariant ones as a subset.
Hence in order to simplify the Abelianization of constraints let
us single out the part of the gauge potentials \( A_{ai} \),
which is invariant under gauge transformations.
Because under a homogeneous gauge transformation the  gauge
potentials transforms homogeneously one can achieve
the separation  of gauge degrees of freedom by the following
simple transformation
\be \label{eq:pcantr}
A_{ai} \left(\bar{Q}, Q^{\ast}\right)
= O_{ak}\left(\bar{Q}\right) Q^{\ast}_{\ ki}~,
\ee
where \( O  \) is an orthogonal matrix, \(  O \in SO(3)\),  and
\( Q^\ast \) is a positive definite symmetric matrix.
This transformation induces a point canonical transformation
linear in the new canonical momenta.
The new canonical momenta \( (P^{\ast}_{\ ik} \ , \bar{P}_{i}) \)
can be obtained using the generating function
\be
{F_4} \left( E; \  \bar{Q}, Q^{\ast}  \right)=
\sum_{a,i}^3 E_{ai} A_{ai} \left(\bar{Q}, Q^{\ast}\right) =
\mbox{tr}
\left(  O ( \bar{Q} ) Q^{\ast} E^T \right) .
\ee
as
\bea
\bar P_j & = & \frac{\partial F_4}{\partial \bar{Q}_j} =
\sum_{a,s,i}^3 E_{ai} \ \frac{\partial O_{as}}{\partial \bar{Q}_j } \
Q^{\ast}_{\ si}  =
\mbox{tr}
\left[ E^T \frac{\partial O}{\partial \bar{Q}_j} \ Q^{\ast}
\right],\\
 P^{\ast}_{\ ik}  & = & \frac{\partial F_4}{\partial Q^{\ast}_{\ ik}} =
\frac{1}{2} \left( O E^T  + E O^T  \right)_{ik}~.
\eea

In order to express the Hamiltonian and the Gauss law constraints in terms
of these new canonical pairs let us write the field strengths $E_{ai}$ in
the form
\be \label{eq:pot}
E_{ai} = O_{ak}\left(\bar{Q}\right) L_{ki} \left(
\bar{P}, P^{\ast}; \bar{Q}, Q^\ast \right)
\ee
with a \( 3 \times 3 \) matrix \( L_{ki} \) to be determined.
One can immediately see that
the symmetric part of the matrix \(L\) is equal to the new momenta
\(P^{\ast}\)
\be
P^{\ast}_{\ ik} = \frac{1}{2} \left( L_{ik} + L_{ki} \right)
\ee
and a straightforward calculation shows that its antisymmetric part is
\be
\frac{1}{2} \left(L_{ik} - L_{ki}\right) =
 \, \epsilon_{ilk} \,( \gamma^{-1})_{ls} \left[
\left(\Omega^{-1} \right)_{sj} \bar P_j - \epsilon_{msn}\left(P^{\ast}
Q^{\ast}\right)_{mn}\right]~,
\ee
with
\be
\Omega_{ij} \, : = \,\frac{1}{2} \, \epsilon_{min}
\left[ \frac{\partial O^T \left(\bar{Q}\right)}{\partial \bar{Q}_j}
\, O ( \bar{Q} ) \right]_{mn}~,
\ee
and
\be
\gamma_{ik} : = Q^{\ast}_{\ ik} -  \delta_{ik} \
\mbox{tr} ( Q^{\ast})~.
\ee
Thus the final expressions for field strengths \( E_{ai} \)  in terms of the
new canonical variables are
\be  \label{eq:potn}
E_{ai} = O ( \bar{Q} )_{ak} \biggl[{\,  P^{\ast}_{\ ki} +
\epsilon _{kli}
 ( \gamma^{-1})_{ls} \left[
\left(\Omega^{-1} \bar P \right)_{s}  - \epsilon_{msn} \left(P^{\ast}
Q^{\ast}\right)_{mn}\,\right]\,
}\biggr]~.
\ee

\subsubsection{Abelianization of first class constraints}

The formulation of the theory in terms of the new variables is adapted
to the procedure of Abelianization.
Using the representations (\ref{eq:pcantr}) and (\ref{eq:potn})
one can easily convince oneself that the
variables \( {Q^\ast}\) and \({P^\ast} \) make no contribution to the
secondary constraints (\ref{eq:phiQstar})  and
\( \bar{Q} \, , \bar{P} \)
enter well-separated from the  physical matter variables
\be
\tilde{\Phi}_a : = O_{as}(\bar{Q}) \Omega^{-1}_{\ sj} \bar{P}_j
-
 \frac{i}{2} \,
{\Pi}^\ast_{\Psi_\alpha} \left(\tau_a \right)_{\alpha \beta}
{Q}^\ast_{\Psi_\beta} = 0~.
\label{eq:4.54}
\ee
In order to deal with the Abelianization it is useful to perform
the following canonical transformation on the Grassmann variables
\bea \label{eq:newgr}
{\Pi}^{\ast}_{\Psi_\alpha} &=: &
{\cal P}^{\ast}_{\Psi_\beta} U_{\beta \alpha } \left( \bar{Q} \right)\, ,\\
{Q}^{\ast}_{\Psi_\alpha}  &=: &  U^{-1}_{\alpha \beta} \left( \bar{Q} \right)
{\cal Q}^\ast_{\Psi_\beta}\, .
\eea
with the unitary matrix \(U\)
in the two dimensional representation of \(SO(3) \) chosen such that
\be
O_{ab} = \frac{1}{2}\mbox{tr} \left(U^+ \tau_a U \tau_b \right)~.
\ee
As a result, the Gauss law constraints (\ref{eq:4.54}) take the form
\be
\tilde{\Phi}'_a : =  \Omega^{-1}_{\ sj} \bar{P}_j
-
\frac{i}{2} \,
{\cal P}^\ast_{\Psi_\alpha} \left(\tau_a \right)_{\alpha \beta}
{\cal Q}^\ast_{\Psi_\beta} = 0\, .
\ee
Hence it is clear that the matrix
\(\Omega^{-1} \) is just the matrix of
Abelianization \(D\) in  (\ref{eq:abelianiz}).
Hence, after performing the Dirac transformation with the matrix
\( D : =  \Omega(\bar{Q}) \) on the constraints
\(\tilde{\Phi}'_a\) the equivalent set of Abelian constraints is
\be
\bar{P}_a  -  \Omega_{\ as} \Theta_s = 0~,
\ee
with
\be
\label{eq:Qstar}
\Theta_a :=
\frac{i}{2} \,
{\cal P}^\ast_{\Psi_\alpha} \left(\tau_a \right)_{\alpha \beta}
{\cal Q}^\ast_{\Psi_\beta}\, .
\ee

\subsection{Reduction due to the Gauss law and the second class constraints }

In the previous section, in accordance with the  general scheme of
reduction formulated in subsection (\ref{subsec:glr}),
the new set of constraints  in canonical form have
been  obtained and the adapted canonical pairs been chosen for
the explicit  implementation of the Gauss laws (\ref{eq:secconstr}) and
the second class constraints (\ref{eq:cansec}).
After having rewritten the model in this form,
the construction of the unconstrained Hamiltonian system
is straightforward. In all expressions we can simply put \(\bar{P} =
\Omega \Theta \) and
\({\bar{\Pi}}_{\Psi_\alpha} = {\bar{Q}}_{\Psi_\alpha} = 0\).
In particular, in terms of the  ``physical'' electric field strength
\( {\cal{E}}_{ai} \)
\be
E_{ai}\ \biggl\vert_{\bar{P} = \Omega \Theta} =: \ O_{ak}(\bar{Q}) \
{\cal{E}}_{ki}
(Q^\ast , P^\ast)~,
\ee
the  physical unconstrained Hamiltonian
\[
H_{phys}:= H_C( P, Q) \,
 \Bigl\vert_{constraint~shell}
\]
may be written as
\be \label{eq:uncYME}
H_{phys}^{D.-Y.-M.} \, = \,
\frac{1}{2}  \mbox{tr}({\cal{E}}^2)
+ \frac{g^2}{4} \left[
\mbox{tr}{}^2 ( {Q^\ast} )^2  -
 \mbox{tr} (Q^\ast)^4
\right]
-  g\,\mbox{tr} \left(j^\ast Q^\ast \right) + i
m \left({\cal P}^\ast_{\Psi_\alpha} \gamma_0 {\cal Q}^\ast_{\Psi_\alpha}
\right) \, .
\ee
where  \( j^\ast \) is the  isospin current in terms of
the new Grassmann variables
\be
\label{eq:jast}
j^\ast_{ia} := \frac{i}{2}
{\cal P}^\ast_{\Psi_\alpha}
 \gamma_i \gamma_0 \left(\tau_a \right)_{\alpha \beta}
{\cal Q}^\ast_{\Psi_\beta}\, .
\ee

With the aid of the identity
\(
~\det\gamma \ \epsilon_{isk} \left( \gamma^{-1}\right)_{sl} =
\epsilon_{alb} \, \gamma_{ia}\,\gamma_{kb}~
\)
and representation (\ref{eq:potn})  for the field strengths,
we find the explicit form for the ``physical'' electric field  strengths
in terms of \( P^\ast \) and \( Q^\ast \)
\be  \label{eq:els}
{\cal{E}}_{ki}(Q^\ast , P^\ast) =
P^\ast_{ik} +  \frac{1}{\det \gamma} \epsilon_{ilk}
\mbox{tr}\, \left(  \gamma {\cal M} \gamma J_l \right)~,
\ee
where \({\cal M}\) denotes the  isospin angular momentum tensor
\bea
{\cal M}_{mn} &: =&  \epsilon_{msn} {\cal J}_s ~.
\eea
Here
\bea
{\cal J}_s &:= & \Theta_s + T_s  \,.
\eea
is the sum of the gauge field isospin vector \( T_s
:= \frac{1}{2} \epsilon_{msn}\left( Q^\ast P^\ast \right)_{mn}
\) and the matter field \(\Theta_s \) defined in (\ref{eq:Qstar}).
With (\ref{eq:els}) the  unconstrained Dirac-Yang-Mills Hamiltonian
reads
\bea \label{eq:uncYMP}
H^{D.-Y.-M.}_{phys} \, & = &\,
\frac{1}{2}  \mbox{tr}(P^{\ast})^2  +
\frac{1}{2 \det^2\gamma }
\mbox{tr}\, \left(\gamma {\cal M}\gamma \right)^2
+ \frac{g^2}{4} \left[
\mbox{tr}{}^2 ({Q^\ast})^2  -   \mbox{tr} (Q^\ast)^4
\right] \nn\\
&-&  g\,\mbox{tr} \left(j^\ast Q^\ast \right) + i
m \left({\cal P}^\ast_{\Psi_\alpha} \gamma_0 {\cal Q}^\ast_{\Psi_\alpha}\right)
\, .
\eea

In order to achieve a more transparent form for the reduced Dirac-Yang-Mills
system (\ref{eq:uncYMP}) one can perform a canonical transformation
expressing the physical coordinates \( Q^\ast \) and \( P^\ast \) in terms
of  new variables adapted for the analysis of the rigid symmetry
possessed by the reduced Hamiltonian system (\ref{eq:uncYMP}).
It is convenient  to decompose the nondegenerate
symmetric matrix \( Q^\ast \) in the following way:
\be
Q^\ast = {\cal R}^{T}(\psi,\theta, \phi)\ {\cal D } \
{\cal{R}}(\psi,\theta, \phi)~,
\ee
with the \( SO(3)\) matrix  \({\cal R}\)
parametrized by the three Euler angles \(\chi_i := (\psi,\theta,\phi )\),
(see Appendix) and with the diagonal  matrix
\(~
{\cal D} : = \mbox{diag}\ ( x_1 , x_2 , x_3 )~.
\)
The corresponding canonical conjugate coordinates \((
p_\psi, p_\theta, p_\phi,  p_i ) \)
can be found by using the generating function
\be
F \left[ x_i, \psi, \theta, \phi; \ P^\ast \right]  : =
 \mbox{tr}\ \left(Q^\ast P^\ast \right) =
 \mbox{tr}\
\left( {\cal R }^T( \chi ) D( x ) {\cal R }( \chi ) P^\ast \right)
\ee
as
\bea
&&
p_{i} = \frac{\partial {F}}{\partial x_i} =
 \ \mbox{tr} \left( P^\ast  {\cal R}^{T} \overline{\alpha}_{i} {\cal R}
\right),\nn\\
&&
p_{\chi_i} = \frac{\partial {F}}{\partial \chi_{i}} =
 \mbox{tr} \left(
\frac{\partial {\cal R}^{T} }{\partial \chi_{i}}
{\cal R} \, \left[ P^\ast Q^\ast- Q^\ast P^\ast\right]
\right),
\eea
where $\overline{\alpha}_{i}$ are the diagonal members of the orthogonal basis
for
symmetric  matrices
\( \alpha_A = ( \overline{\alpha}_i ,\ \alpha_i ) \  i = 1, 2, 3 \) \
given explicitly in the Appendix.
The original physical momenta  \( P^\ast_{ik} \) can then be expressed
in terms of the new canonical variables  as
\bea \label{eq:newmom}
{P}^\ast  =
{\cal R }^T \ \left(
         \sum_{s=1}^3  p_s \, \overline{\alpha}_{s} +
 \sum_{s=1}^3 {\cal P}_s \, {\alpha}_{s}\right)
\ {\cal R }
\,
\eea
with
\bea
{\cal P}_{i}  : =  {\frac{\xi_i}{ x_j - x_k }}, \,\,\,\,\,
(cyclic\,\,\,\, permutation \,\,\, i\not=j\not= k )
\eea
and  the $SO(3)$ left-invariant Killing vectors
\bea
&& \xi_1  : =
\frac{\sin\psi}{\sin\theta}\ p_\phi +
\cos\psi \  p_\theta - \sin\psi \cot\theta \ p_\psi~,\\
&& \xi_2 : =
-\frac{\cos\psi}{\sin\theta}\ p_\phi +
\sin\psi \  p_\theta + \cos\psi \cot\theta \ p_\psi~,   \\
&& \xi_3  : = p_\psi~.
\eea
Representing the physical electric field strengths
${\cal E}_{ai}$ in the alternative form
\be   \label{eq:nels}
{\cal{E}}_{ik} = {P^\ast}_{ik} +
\frac{1}{\det \gamma}\,  \left( \gamma J_s \gamma \right)_{ik} \,{\cal J}_s~,
\ee
with the SO(3) generators $J_s$ given in explicit form in the Appendix,
we finally get the following physical Hamiltonian defined on
the unconstrained phase space
\bea \label{eq:DYM}
H^{D.-Y.-M.}_{phys} &=& \frac{1}{2} \sum_{s=1}^3  p^2_s +
\frac{1}{4} \sum_{s=1}^3  {\cal P}^2_s  +
\frac{1}{4}\sum_{cyclic}
\left(
\frac{\xi_i + \Theta_i  }{ x_j + x_k }\right)^2 +
\frac{g^2}{2} \  \sum_{i < j}  x_i^2 x_j^2  \nn\\
&-& g  \sum_{s=1}^3 j_{ss}^\ast x_s
+ i m \left( {\cal P}^\ast_{\Psi_\alpha} \gamma_0 {\cal Q}^\ast_{\Psi_\alpha}
\right)~.
\eea
Note that for the pure Yang-Mills system (\ref{eq:DYM}) reduces to
\be  \label{eq:PYM}
H^{Y.-M.}_{phys} = \frac{1}{2} \sum_{s=1}^3  p^2_s +
\frac{1}{2}\sum_{cyclic}  \xi^2_i \frac{x_j^2 +  x_k^2}{\left(x_j^2 -
x_k^2\right)^2} +
     \frac{g^2}{2} \  \sum_{i < j}  x_i^2 x_j^2~.
\ee
This completes our reduction of the spatially homogeneous constraint
Dirac-Yang-Mills system to the equivalent unconstrained system
describing the dynamics of the physical dynamical degrees of
freedom.
However, apart from this reduction due to the underlying gauge symmetry,
there is the possibility to realize another type of
reduction connected with the rigid symmetry
admitted by the unconstrained system (\ref{eq:uncYMP}).
For simplicity,  the discussion in the next section
will be restricted to the pure Yang-Mills system and we shall
 show how to further reduce the obtained
\(12\)-dimensional system (\ref{eq:PYM})
to an \( 8 \)-dimensional one
in general and, for a special case to
a \( 6 \)-dimensional one using the corresponding first integrals.

\subsection{Further reduction using first integrals}

The  reduced Yang-Mills theory (\ref{eq:PYM})  has a rigid  symmetry
connected with the existence of the first integrals
\be \label{eq:firstint}
I_i = \epsilon_{ijk} E_{aj}A_{ak}~.
\ee
For the  subsequent reduction in the number of degrees of freedom
we shall use the integrals of motion (\ref{eq:firstint}).
One can verify that in terms of the new variables they read
\(~
I_i = {\cal R}_{ik} \xi_k~.
\)
In contrast to the reduction due to first class constraints
the values of first integrals are arbitrary and
depend on the initial conditions.
In this case reduction means to consider the subspaces of phase space
which are the levels of fixed values for these first integrals
\be \label{eq:msur}
I_i = c_i~,
\ee
and the subsequent construction of the quotient space
with respect to the rigid symmetry group.
Therefore, in contrast to the reduction that we have done
before, the first integrals in general are a mixed system of
first and second class constraints.
By ``in general'' we mean that not all constants of motion  \( c_i\) are zero.
In this case the rank of the Poisson matrix is
\(~
\mbox{rank}~||\{ I_i, I_k \}|| \bigl\vert_{I_i = c_i}= 1~,
\)
which means that there is one first class constraint and one
pair of second class constraints
\footnote{In the exceptional case when \( c_i = 0 \) we
can consider the three integrals as first class constraints,
and this circumstance
leads to a further reduction of our system. }.
The problem is now to separate the algebra of constraints and to find the
equivalent set of constraints
\(~
\Psi_i  = 0~,
\)
so that the first class constraint \( \Psi_1 \)
forms the center of the algebra
\be
\{\Psi_1 , \Psi_i\} = 0
\ee
and the pair of second class constraints
obey the canonical algebra
\be
\{\Psi_2 , \Psi_3\} = 1~.
\ee
After having passed to new variables in the last section in order to isolate
the gauge degrees of freedom from the physical ones, we shall now
perform another canonical transformation from the physical variables
to new physical variables so that one of the new momenta coincides with
the first class constraint \(\Psi_1 \)
and another pair of new canonical variables coincides with the pair of
second class constraints
\(\Psi_2 \) and \( \Psi_3\).
In terms of these new canonical variables the reduced
system is obtained by
reducing the Hamiltonian (\ref{eq:uncYMP}) to the
integral surface  (\ref{eq:msur}).
As result the new Hamiltonian will
depend on \( 4 \) canonical pairs and
one parameter which reflects the existence of the integrals of motion.
To demonstrate this let us choose the
integral constants as \( c_i = (0, 0, c) \) without  loss of generality.
One can then write down the needed new set of
constraints, describing the surface (\ref{eq:msur}), in the form
\bea
&&\Psi_1 := I_1^2 +I_2^2 +I_3^2 -c^2 = 0~, \nn\\
&&\Psi_2 := \arctan\left( \frac{I_2}{I_3} \right) = 0~, \\
&&\Psi_3 := {I_1} = 0~. \nn
\eea
We are now ready to perform the transformation to special canonical
variables so that the pair of second class constraints
is equal to the one pair of the canonical variables and one
equal to the new momentum
\footnote{
This type of variables are well-known
from rigid body theory as Depri \cite{Arkhangel} or
Andoyer \cite{Andoyer,Arnold} variables used in celestial mechanics.}
\be
\Pi_0 : = \Psi_1~, \,\,\,\, \Pi_1 := \Psi_3~, \,\,\,\,
X_1 := \Psi_2~.
\ee
and complete the set of canonical coordinates by
the following pair
\be
X_2:= \arctan\left( \frac{\xi_1}{\xi_2} \right)~, \,\,\,\,
\Pi_2: = p_\psi~.
\ee
The canonical conjugate coordinate \(X_0\) can be determined
with the help of the generating function
\be
 F[\psi, \theta, \phi, \Pi_i] = \Pi_1\phi + \Pi_2 \psi +
\int^\theta \frac{d\alpha}{\sin\alpha}
\sqrt{\Pi_0^2 \ \sin^2\alpha - \Pi_1^2 -\Pi_2^2 +2 \Pi_1 \Pi_2 \cos\alpha}~.
\ee
Due to the  symmetry  the
\(X_0\) is a cyclic coordinate and the reduced Hamiltonian depends
only on the canonical pair
\[
\Pi_2 = p_\psi\,, \,\,\,\,\, X_2:= \arctan\left( \frac{\xi_1}{\xi_2} \right) \
\Bigl\vert_{I_i=c_i} = \psi~.
\]
Hence, using the first integrals (\ref{eq:firstint}),
the pure Yang-Mills Hamiltonian (\ref{eq:PYM})
can be further reduced to
\bea \label{eq:PYMr}
H^{Y.-M.~\ast}_{phys} &=& \frac{1}{2} \sum_{s=1}^3  p^2_s +
\frac{1}{2} p_\psi^2 \left[
\frac{x_1^2 +  x_2^2}{\left(x_1^2 - x_2^2\right)^2} -
\sin^2\psi\frac{x_2^2 +  x_3^2}{\left(x_2^2 - x_3^2\right)^2}
   - \cos^2\psi\frac{x_3^2 +  x_1^2}{\left(x_3^2 - x_1^2\right)^2}
\right] \nn\\
&+&
     \frac{g^2}{2} \  \sum_{i < j}  x_i^2 x_j^2 + V_C~,
\eea
where in accordance  with the general scheme of reduction there
arises the additional so-called reduced potential term
\be
V_C :=
c^2 \left[\sin^2\psi\frac{x_2^2 +  x_3^2}{\left(x_2^2 - x_3^2\right)^2} +
\cos^2\psi\frac{x_3^2 +  x_1^2}{\left(x_3^2 - x_1^2\right)^2}
\right]~.
\ee
It is interesting to point out the difference
between the reduced Yang-Mills Hamiltonian
(\ref{eq:PYM}) and the corresponding one in the
recent work by B. Dahmen and  B. Raabe.
In contrast with their representation for the gauge potentials, in which
the gauge degrees of freedom are mixed with the rigid rotational
cyclic coordinates, we have started with the explicit separation
of all  physical degrees, including the  rotational ones.
And only after the reduction in the number of
 degrees of freedom due to the rigid symmetry the
obtained Hamiltonian (\ref{eq:PYMr}) coincides with the one
obtained in the work by B. Dahmen and  B. Raabe
\cite{DahRaab} for pure Yang-Mills mechanics.

\section{Concluding remarks}

As mentioned in the introduction our investigation
has pursued two goals. One is pure theoretical interest.
Due to the homogeneity condition SU(2) Dirac-Yang-Mills field theory has
greatly simplified to a finite dimensional mechanical system, for which
one can describe
the equivalent unconstrained system in an explicit way.
However, apart from this reason, there is also an interesting application
of this model.
It has been known for a long time, that, if one considers the Euclidean
QCD effective action  as a function of the non-abelian
electric and  magnetic fields  \(E \) and \(B \), one finds that
there are field configurations,  corresponding to
nonvanishing \(E\) and \(B\) fields,
for which the value of the effective action is lower than that for
\(E=0\) and \(B=0\) \cite{Savv}.
This observation indicates a drastic difference between the
true ground state of QCD and the corresponding
perturbative vacuum and constitutes the basis of all models of
condensates.
One of the main reasons to study the dynamics of spatially
constant Yang-Mills fields, is the faith
that the corresponding zero momentum quantum operators are very important
for the description of the QCD ground state due to the presence of the
IR singularity.
There are many attempts to
exploit the homogeneity approximation for gluon fields with
the aim to shed light on the vacuum structure of QCD.
We also adhere to this position and our task in this note was to prepare
the classical description of  Yang-Mills mechanics in a form that we are
going to exploit for the description of squeezed vacuum \cite{squeez}.


\acknowledgments

We are grateful for discussions with  Profs.  G.Lavrelashvili,
 V.P. Pavlov,
V.N. Pervushin,  A.N.Tavkhelidze.
One of us (A.K.) would like to thank Prof. G.R\"opke
for kind hospitality at the MPG AG ''Theoretische Vielteilchenphysik''
Rostock where part of this work has been done
and the Max-Planck Gesellschaft for providing a
stipendium during the visit.
This work was supported also by the Russian Foundation for
Basic Research under
grant No. 96-01-01223 and by the Heisenberg-Landau program.
H.-P. P.  acknowledges support by the Deutsche Forschungsgemeinschaft
under grant No. Ro 905/11-1.

\appendix

\section{Notations and some formulae}

\subsection{Definition of configuration variables}

\(SU(2)\) Dirac-Yang-Mills theory considered in this paper includes
as dynamical variables the set of spin-1 gauge fields \( A_\mu :=
\, A^a_\mu {\tau}_a/2 ,~a = 1,2,3 \) in the adjoint representation of
SU(2), with the corresponding field strengths
\bea
F_{\mu \nu} & : = & F^a_{\mu\nu} \ \tau^a/2,~ \\
F^a_{\mu\nu}       & : = & \partial_\mu A_\nu^a  -  \partial_\nu A_\mu^a
+ g \epsilon^{abc} A_\mu^b A_\nu^c~,
\eea
and the matter spinor (Dirac conjugate spinor) field variables
 \( \Psi (\bar{\Psi})\) in the fundamental representation of SU(2) with values
\( \Psi_\alpha := ( \Psi_\alpha^1, \dots , \Psi_\alpha^4) \)
obeying the Grassmann  algebra
\be
\Psi_\alpha^i \Psi_\beta^j + \Psi_\beta^j \Psi_\alpha^i  =0~.
\ee

\subsection{Hamiltonian structures}

Generalized Poisson brackets for functions
on a phase space spanned by both
even and odd coordinates
\( Z_A:= ((Y, P_Y), (A_i, E_i); (\Psi_\alpha , P_{\psi_\alpha} ))\)
are defined as
\be
\{ F(Z), G(Z)\} :=  \sum_{A, B} F
\frac{\stackrel{\leftarrow}{\partial}}{\partial Z_A}
\omega_{AB}
\frac{\stackrel{\rightarrow}{\partial}}{\partial Z_B} G~.
 \ee
The nonvanishing components of the canonical symplectic form
\(
\omega_{AB}:= \{ Z_A, Z_B \}
\)
read explicitly
\be
\{ Y_{a}, P_Y^{b} \} =  \delta_a^b ~, \quad \quad \quad
\{ A_{ai}, E^{bj} \} = \delta_i^j \delta_a^b
\ee
for bosonic degrees of freedom
\bea
\{ \Psi_{\alpha}, P_{\Psi_{\beta}} \} =
\{ P_{\Psi_{\beta}}, \Psi_{\alpha} \} = - \, \delta_{\alpha \beta}~, \nn\\
\{ \bar{\Psi}_{\alpha}, P_{\bar{\Psi}_{\beta}} \} =
\{ P_{\bar{\Psi}_{\beta}}, \bar{\Psi}_{\alpha} \} =
- \, \delta_{\alpha \beta}
\eea
for fermionic degrees of freedom.

\subsection{The Euler parametrization for S0(3) group}

The conventional representation of \(SO(3)\) group
elements  in terms of Euler angles
\be
 {\cal R }(\psi,\theta, \phi) =
e^{\psi J_3}e^{\theta J_1}e^{\phi J_3}
\ee
has been used in main text with the following matrix realization for
the generators \( J_i \) obeying the  \( SO(3) \)  algebra
\ \( [ J_i, J_j] = \epsilon_{ijk} \ J_k\)
\be
{J}_{1} = \left (
\begin{array}{ccc}
0      &   0      &     0      \\
0      &   0      &   - 1      \\
0      &   1      &     0
\end{array}
\right ),
\quad \quad
{J}_{2} = \left (
\begin{array}{ccc}
0      &   0      &   - 1     \\
0      &   0      &     0     \\
1      &   0      &     0
\end{array}
\right ),
\quad \quad
{J}_{3} = \left (
\begin{array}{ccc}
0      &  - 1      &    0     \\
1      &    0      &    0     \\
0      &    0      &    0
\end{array}
\right )~.
\ee

\subsection{Basis for symmetric matrices}

We use the orthogonal basis
$\alpha_A = ( \overline{\alpha}_i , \ \alpha^i )$
for symmetric matrices. They read explicitly
\be
\overline{\alpha}_1 = \left (
\begin{array}{ccc}
1      &   0      &    0     \\
0      &   0      &    0     \\
0      &   0      &    0
\end{array}
\right )~,
\quad \quad
\overline{\alpha}_2 = \left (
\begin{array}{ccc}
0      &   0      &    0     \\
0      &   1      &    0     \\
0      &   0      &    0
\end{array}
\right )~,
\quad \quad
\overline{\alpha}_3 = \left (
\begin{array}{ccc}
0      &   0      &    0     \\
0      &   0       &    0     \\
0      &   0      &    1
\end{array}
\right )~,
\ee

\be
{\alpha}^{1} = \left (
\begin{array}{ccc}
0      &   0      &    0     \\
0      &   0      &    1     \\
0      &   1      &   0
\end{array}
\right )~,
\quad \quad
{\alpha}^{2} = \left (
\begin{array}{ccc}
0      &   0      &    1     \\
0      &   0      &    0     \\
1      &   0      &   0
\end{array}
\right )~,
\quad \quad
{\alpha}^{3} = \left (
\begin{array}{ccc}
0      &   1      &    0     \\
1      &   0      &    0     \\
0      &   0      &   0
\end{array}
\right )~.
\ee
They obey the following orthonormality relations:
\be
\mbox{tr}\ (\overline{\alpha}_i \overline{\alpha}_j) = \delta_{ij}~,
\quad \quad
\mbox{tr}\ ({\alpha}_i {\alpha}_j) = 2 \delta_{ij}~,
\quad \quad
\mbox{tr}\ (\overline{\alpha}_i {\alpha}_j) = 0~.
\ee


\begin{thebibliography}{99}
\bibitem{DiracL}
P.A.M. Dirac,
{\it Lectures on Quantum Mechanics\/}, Belfer Graduate School of Science,
(Yeshive University Press, New York, 1964).
%
\bibitem{KonPop}
N.P. Konopleva, V.N. Popov, {\it Gauge fields },
( Atomizdat, Moscow, 1972) (in Russian).
%
\bibitem{Sunder}
K. Sundermeyer, {\it Constrained Dynamics\/},
Lecture Notes in Physics N 169,
(Springer Verlag, Berlin - Heidelberg - New York, 1982).
%
\bibitem{GitTyut}
D.M. Gitman, I.V. Tyutin,
{\it Quantization of Fields With Constraints  \/},
(Springer Verlag, Bonn, 1990 ).
%
\bibitem{HenTeit}
M. Henneaux, C. Teitelboim,
{\it Quantization of Gauge Systems \/},
(Princeton University Press, Princeton, NJ, 1992).
%
\bibitem{YM}
C.N. Yang, R.L. Mills,  Phys. Rev {\bf 96}, 191 (1954).
%
\bibitem{Geom}
V.N. Gribov, Nucl. Phys. B {\bf 139}, 1 (1978).\\
I.M. Singer, Comm. Math. Phys. {\bf 60}, 7 (1978);
Physica Scripta {\bf 24},  817 (1981).\\
O. Babelon, C.M. Viallet, Comm. Math. Phys. {\bf 81}, 515 (1981).\\
M.S. Narasimhan, T.R. Ramadas, Comm. Math. Phys. {\bf 67}, 21 (1979).\\
P.K. Mitter, in: Recent Developments in Gauge
Theories ed. t'Hooft G., (Plenum Press, New-York,  1980).\\
M.F. Atyah, {\it Geometry of Yang-Mills Fields},
(Pisa: Scuola Normale Superiore,  1979).\\
V. Moncrief, in:  Springer Lecture Notes in Mathematics
{\bf 836}, 276 (1980).
%
\bibitem{GoldJack}
J. Goldstone, R. Jackiw, Phys. Lett. B {\bf 74\/}, 81 (1978).
%
\bibitem{Baluni}
V. Baluni, B. Grossman, Phys. Lett B {\bf  78\/}, 226 (1978).
%
\bibitem{Faddeev79}
A.G. Izergin, V.F. Korepin, M.E. Semenov - Tyan - Shanskii,
L.D. Faddeev, Teor. Mat. Fiz. {\bf 38 }, 3 (1979).
%
\bibitem{DasKakuTown}
A. Das, M. Kaku, P.K. Townsend, Nucl.Phys B {\bf 149}, 109 (1979).
%
\bibitem{ChrLee}
N.H. Christ, T.D. Lee,  Phys. Rev. {\bf 22 \/}, 939 (1980).
%
\bibitem{Pervush1}
V.N. Pervushin, Teor. Mat. Fiz. {\bf 45}, 327 (1980).
%
\bibitem{Simon}
Yu. Simonov, Sov. J. Nucl. Phys. {\bf 41}, 835 (1985).
%
\bibitem{Tav}
V.V. Vlasov, V.A. Matveev, A.N. Tavkhelidze, S.Yu. Khlebnikov,
M.E. Shaposhnikov, Phys. of Elem. Part. Nucl. {\bf 18}, 5 (1987).
%
\bibitem{Newman2}
E.T. Newman, C. Rovelli, Phys. Rev. Lett {\bf 69}, 1300 (1992).
%
\bibitem{KJohnson}
D. Freedmann, P. Haagensen, K. Johnson, J. Latorre,
MIT Preprint CTP 2238 (1993).
%
\bibitem{DMacmul}
M. Lavelle, D. McMullan, Phys. Rep. {\bf 279}, 1 (1997).
%
\bibitem{KhZer0}
A.M. Khvedelidze, V.N. Pervushin, Helv. Phys. Acta. {\bf 67}, N6, 637 (1994).
%
\bibitem{MatSav}
S.G. Matinyan, G.K. Savvidy and N.G. Ter-Arutyunyan-Savvidy,
Sov.Phys. JETP {\bf 53}, 421 (1981).
%
\bibitem{AsatSav}
H.M. Asatryan and G.K. Savvidy, Phys.Lett. A {\bf 99 }, 290 (1983).
%
\bibitem{MSolov}
M.A. Soloviev, Teor. Mat. Fyz. {\bf 73}, 3 (1987).
%
\bibitem{Gotay}
M.J. Gotay, J.Geom. Phys. {\bf 6}, 349 (1989).
%
\bibitem{DahRaab}
B. Dahmen, B. Raabe, Nucl. Phys. B {\bf 384}, 352 (1992).
%
\bibitem{Shanmugadhasan}
S. Shanmugadhasan, J. Math. Phys {\bf  14\/}, 677 (1973).
%
\bibitem{Cartan}
E.Cartan, {\it Lecons sur les invariant integraux}, (Hermann, Paris, 1922).
%
\bibitem{Marsd}
J. Marsden, A.Weinstein, Rep. Math. Phys. {\bf  5\/}, 121 (1974).
%
\bibitem{Olver}
P. Olver, {\it Applications of Lie Groups to Differential Equations\/},
Graduate Text in Mathematics,
(Springer Verlag, New York-Berlin - Heidelberg - Tokyo, 1986).
%
\bibitem{Arnold}
V.I. Arnold, V.V. Kozlov, A.I. Neishtadt,
{\it Mathematical Aspects of  Classical and Celestial
Mechanics \/}, in: Dynamical Systems III,
(Springer Verlag, New York-Berlin,  1988).
%
\bibitem{Perelomov}
A.M. Perelomov, {\it Integrable Systems in Classical Mechanics
and Lie's Algebra}, (Nauka, Moscow, 1990) (in Russian).
%
\bibitem{Dirac49}
P.A.M. Dirac,  Rev. Mod. Phys. {\bf 21 \/}, 392 (1949).
%
\bibitem{Faddeev69}
L.D. Faddeev, Theor. Math. Phys. {\bf 1}, 1 (1969).
%
\bibitem{Dirac59}
P.A.M. Dirac, Phys. Rev. {\bf 114}, 924 (1959).
%
\bibitem{Whittaker}
E.T. Whittaker, { \it A Treatise on the Analitical Dynamics
of Particles and Rigid bodies\/}, (Cambridge University Press,
Cambridge, 1937).
%
\bibitem{BatVil}
I.A. Batalin, G.A. Vilkovisky, Nucl. Phys B {\bf 234}, 106 (1984).
%
\bibitem{GKP}
S.A. Gogilidze, A.M. Khvedelidze, V.N. Pervushin
Phys. Rev. D {\bf 53}, 2160 (1996).
%
\bibitem{GKHAbel} S.A. Gogilidze, A.M. Khvedelidze, V.N. Pervushin
J. Math. Phys. {\bf 37}, 1760 (1996).
%
\bibitem{Bergman}
P.G. Bergman and I. Goldberg, Phys. Rev. {\bf  98 \/}, 531 (1955).
%
\bibitem{Prokh}
L.V. Prokhorov, S.V. Shabanov, Usp. Fiz. Nauk. {\bf 161}, N2, 13 (1991).
%
\bibitem{Savv}
G.K. Savvidy, Phys. Lett. B {\bf 71 }, 133 (1977).
%
\bibitem{Arkhangel}
Y. A. Arkhangelskii, {\it Analitical  dynamics of rigid body},
(Nauka, Moscow, 1977) (in Russian).
%
\bibitem{Andoyer}
H.Andoyer, {\it Cours de m\`ecanique c\`eleste}
v 1., (Gauthier-Villars, Paris, 1923)
%
\bibitem{squeez}
D. Blaschke, H.-P. Pavel, V.N. Pervushin, G. R\"opke
                 and M.K. Volkov, Phys. Lett. B {\bf 397}, 129 (1997);
{\it Squeezed gluon condensate and the mass of the $\eta'$}, hep-ph/9706528.
\end{thebibliography}
\end{document}